# SYNTHESIS OF SmFeAsO BY AN EASY AND VERSATILE ROUTE AND ITS PHYSICAL PROPERTY CHARACTERIZATION

V.P.S. Awana*, Anand Pal, Arpita Vajpayee and H. Kishan
National Physical Laboratory, Dr. K.S. Krishnan Marg, New Delhi 110012, India

G. A. Alvarez[1]
Institute for superconducting and Electronic Materials, Univ. of Wollongong, NSW 2522, Australia

K. Yamaura, and E. Takayama-Muromachi
Advanced nano-materials Laboratory, National Institute for Material Science (NIMS), Tsukuba, Ibaraki 305-0044, Japan

**Abstract**

We report synthesis, structure, electrical transport and heat capacity of SmFeAsO. The title compound is synthesized by one-step encapsulation of stoichiometric FeAs, Sm, and $Sm_2O_3$ in an evacuated ($10^{-5}$ Torr) quartz tube by prolong (72 hours) annealing at 1100°C. The as synthesized compound is crystallized in tetragonal structure with P4/nmm space group having lattice parameters $a$ = 3.93726(33) A and $c$ = 8.49802(07) A. The resistance ($R$-$T$) measurements on the compound exhibited ground state spin-density-wave (SDW)-like metallic steps below 140 K. Heat capacity $C_P(T)$ measurements on the title compound, showed an anomaly at around 140 K, which is reminiscent of the SDW ordering of the compound. At lower temperatures the $C_P(T)$ shows a clear peak at around 4.5 K. At lower temperature below 20 K, $C_p(T)$ is also measured under an applied field of 7 Tesla. It is concluded that the $C_P(T)$ peak at 4.5 K is due to the anti-ferromagnetic(AFM) ordering of $Sm^{3+}$ spins. These results are in confirmation with ordering of Sm in $Sm_{2-x}Ce_xCuO_4$.

PACS: 64.62Bf; 74.70Ad,; 74.81 Bd

Key words: A. High $T_c$ materials; A. Magnetically ordered materials; D. Superconductivity

[1] Presenting Author;

*: Corresponding Author

Dr. V.P.S. Awana

National Physical Laboratory, Dr. K.S. Krishnan marg, New Delhi-110012, India

Fax No. 0091-11-45609310: Phone No. 0091-11-45608329

e-mail-awana@mail.nplindia.ernet.in: www.freewebs.com/vpsawana/

## 1. Introduction

Very recent invention of superconductivity of up to 58 K in REFeAsO (RE = La, Pr, Sm, Nd, Gd) has been of tremendous interest to the scientific community [1-10]. Besides the Cu based high $T_c$ superconductors (HTSc) [11], the REFeAsO is the only known class of superconductors till date, having their superconducting transition temperature ($T_c$) outside the so-called strong BCS (Bardeen Cooper and Schreifer) limit i.e. 40 K. Further, the Fe based compound provides an excellent opportunity to the theoreticians to think out side the cuprate families (HTSc) in search for the mechanism of high $T_c$ superconductivity [12-15]. Unlike the HTSC, the synthesis route for the REFeAsO superconductors is quite complicated. Basically this is the reason that even after 4-6 months of their invention the gold rush is not like HTSc. In late 1986 and early 1987-88, the reporting of $T_c$ was on daily basis like newspaper updates. Though the cond-mat arxiv preprint cite brings about 5 articles on daily basis in last two-three months, the material is yet confined mainly to couple of Japanese, Chinese and some USA and Europe laboratories. Infact, the REFeAsO class of compounds is mainly synthesized by the high-pressure high-temperature (HPHT) process with pressure as high as 4-6 GPa at 1250$^o$C [4-6]. Many researchers do not have the HPHT synthesis technique available in their laboratories. Some reports are available for the normal pressure synthesis route as well [1-3,7-10]. The key step for normal pressure synthesis is to encapsulate the sample in evacuated/Ar filled quartz tube and annealing up to 1150$^o$C. One of the difficulties in this process is the stability of quartz at high temperature (1150$^o$C). In particular, when carriers are doped via fluorine (F) doping, the quartz tube corrodes and resulting breaking of vacuum. In the current communication, we report synthesis of SmFeAsO in an evacuated ($10^{-5}$ Torr) quartz tube by prolong (72 hours) annealing at 1100$^o$C. The resultant compounds are found to be nearly single phase but not superconducting due to lack of carriers. The resistivity and heat capacity measurements established the SDW (spin density wave) type ground state of the compound below 140 K. Further, the $Sm^{3+}$ spins order AFM below 4.5 K in this compound.

## 2. Experimental

Stoichiometric amounts of FeAs, Sm, and $Sm_2O_3$ are weighed and mixed thoroughly in formula ratio SmFeAsO. The weighed and mixed palletized powders are encapsulated in SS (stainless steel) tubes having small pinholes at both the ends. These are further encapsulated in an evacuated ($10^{-5}$ Torr) quartz tube and annealed at 1100$^o$C for over 72 hours respectively in a single step, followed by cooling to room temperature over a span of 12 hours. This procedure is more or less similar to that as reported earlier [1,3,16]. The X-ray diffraction pattern of the compound was taken on Rigaku mini-flex II diffractometer. The resistivity measurement was carried out by conventional four-probe method on a close cycle refrigerator in temperature range of 12 to 300 K. Heat capacity measurements were carried out on Quantum design PPMS.

### 3. Results and Discussion

The room temperature powdered sample's X-ray diffraction (XRD) pattern and its Rietveld analysis are shown in lower inset of Fig.l. The XRD pattern of the compound is fitted on the basis of tetragonal, *P*4/*nmm* space group. It indicates that besides the majority phase (tetragonal *P*4/*nmm*) few weak impurity lines either from Fe, As, FeAs, or SmAs are also seen in the XRD pattern. The refined structural parameters of the studied sample are listed in Table 1. The lattice parameters are found to be: $a$ = 3.93726(33) A and $c$ = 8.49802(07) A. These values of lattice parameters are in confirmation with earlier reports [3,17,18,19]. The lower inset of Fig. 1 shows the reasonably good fitting of the observed pattern of XRD.

Resistance versus temperature (*R*-*T*) plot of the SmAsFeO compound is shown in upper inset of Fig.1. The *R*-*T* behavior is less metallic comparatively with a relatively high resistance around 24 mΩ at room temperature, which decreases slightly as the temperature goes down to around 140 K. At 140 K, the resistance drops abruptly. Below 140 K the SmAsFeO exhibits a metallic step down to the lowest measured temperature of 20 K. The sharp metallic step below 140 K is in agreement with the earlier reported results [1-10] and it arises due to the spin density wave (SDW) magnetic structure of the compound.

Heat capacity ($C_P$) versus temperature plot for the presently studied SmAsFeO compound is shown in main panel of Fig.1. The absolute value of $C_P$ at 200 K is around 90 J/mol-K, which is though comparable but slightly less than the only other available report on $C_P(T)$ of SmAsFeO compound [20]. With decrease in $T$ the $C_P$ goes down continuously, in a similar fashion as in ref.20. A distinct hump/kink is seen in $C_P(T)$ at around 140 K. Interestingly this is the same temperature, where the metallic step is observed in $R(T)$. The metallic step is reminiscent of the SDW magnetic anomaly in the system. It is also known, that besides SDW the ground state non-superconducting REFeAsO systems go through a structural phase transition simultaneously. With further decrease in temperature the $C_p(T)$ goes down before exhibiting a sharp peak at around 4.5 K. The sharp peak at around 4.5 K is seen earlier in ref.20 as well and is assigned to the AFM ordering of $Sm^{3+}$ spins. In brief the $C_p(T)$ measurements exhibited a shallow hump at around 140 K and a sharp peak at 4.5 K. The former is probably due to the SDW magnetic ordering and the structural phase transition and the later from AFM ordering of $Sm^{3+}$ spins. To understand the nature of the observed two characteristic $C_P(T)$ peaks i.e., $C_P$(140K) and $C_P$(4.5K), the same are zoomed and shown in Fig. 2(a) and (b) respectively. It is noted that sharp specific heat peak obtained at 4.5 K at zero magnetic field does shift anywhere on temperature scale in the application of 7 Tesla magnetic field.

We have separated the magnetic and non-magnetic contributions to $C_p$ by fitting of data in temperature range 12 K $\leq T \leq$ 20 K using the equation [20,21]

$$C_p^{NM}(T) = \gamma T + \beta T^3 \qquad (1)$$

where the linear term corresponds to the electronic contribution and cubic term corresponds to lattice contribution to the specific heat. In Fig. 2(c), $C_p / T$ versus $T^2$ is plotted for SmFeASO in zero magnetic field together with the fitting values (solid line). The values of $\gamma$ and $\beta$ are found to be 48.79 mJ/mole $K^2$ and 0.24 mJ/mole $K^4$ respectively. This value of $\gamma$ is lower than that obtained by ding et al in Ref. 20. The contribution of magnetic correlation to the measured $C_P(T)$ is calculated by using equation [20,21]

$$C_p^{mag}(T) = C_p(T) - C_p^{NM}(T) \qquad (2)$$

and is plotted in Fig. 2(d). The temperature dependence of entropy associated with the magnetic transition is calculated from the $C_p^{mag}(T)$ and shown in inset of Fig. 2(d). The magnetic entropy saturates at temperature above 4.5 K to a value ≈ 4.5 J/mole K. The usual entropy value of a doublet ground state of $Sm^{+3}$ is 1.85$R$ ln 2, where $R$ is the gas constant. The estimated entropy value is obviously lower than the expected one of a doublet ground state. The reduced entropy can be explained by the fact that itinerant electrons with heavy effective mass are involved in the transition [21].

In summary, the specific heat of non-superconducting SmFeAsO compound has been studied systematically. First, a jump in specific heat value is observed at 140 K, which is attributed to structural or SDW transition. Again a sharp peak is obtained at $T$ = 4.5 K corresponding to the anti-ferromagnetic ordering of $Sm^{+3}$ ions in the system.

Table 1. Rietveld refined structure parameters of SmFeAsO. Space group: *P*4/*nmm*

| Atom | Site | x | y | z | Occupancy |
|---|---|---|---|---|---|
| Sm | 2c | 0.25 | 0.25 | 0.1356(8) | 92.5(9) |
| Fe | 2b | 0.75 | 0.25 | 0.5 | 100.1(4) |
| As | 2c | 0.25 | 0.25 | 0.6514(15) | 94.3(8) |
| O | 2a | 0.75 | 0.25 | 0 | 95.3(6) |

$a$ = 3.93726(33) A and $c$ = 8.49802(07) A

*Rp*: 3.26%, *Rwp*: 4.20%, *Rexp*: 2.47%, $\chi^2$: 2.90

**Figure Captions**

Figure 1: Specific heat of SmFeAsO sample (main panel); Observed and fitted XRD pattern (inset 1) and Resistivity vs temperature plot (inset 2).

Figure 2(a): Enlarged view of specific heat of SmFeAsO in the high temperature range.

Figure 2(b): Enlarged view of specific heat of SmFeAsO in the low temperature range at $H = 0$ and 7 T.

Figure 2(c): $C_p/T$ vs $T^2$ plot for SmFeASO in $H = 0$ T and its linear fitting.

Figure 2(d): Magnetic specific heat of SmFeAsO and entropy associated with the magnetic transition (in inset).

Figure 1

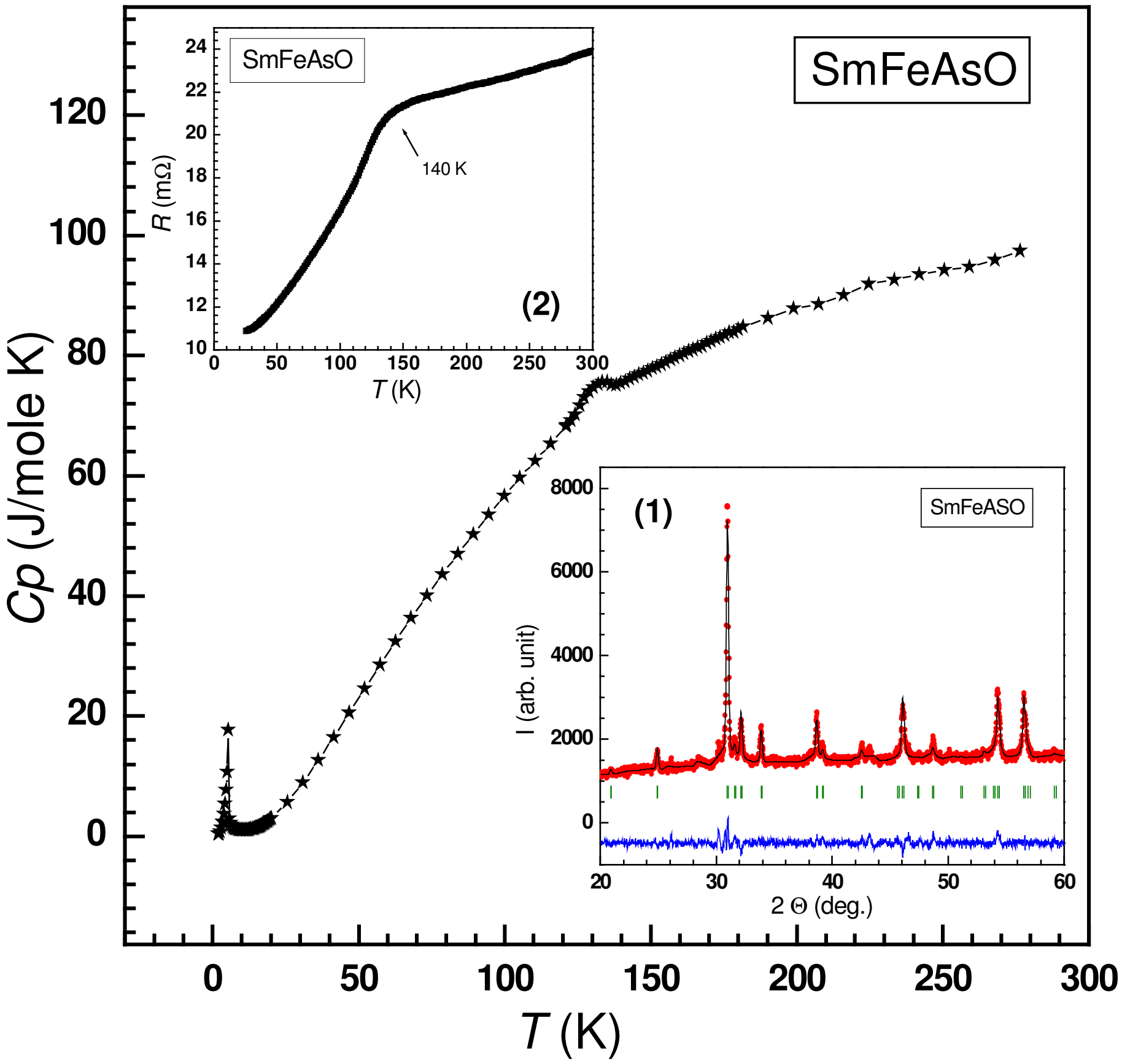

Figure 2 (a)

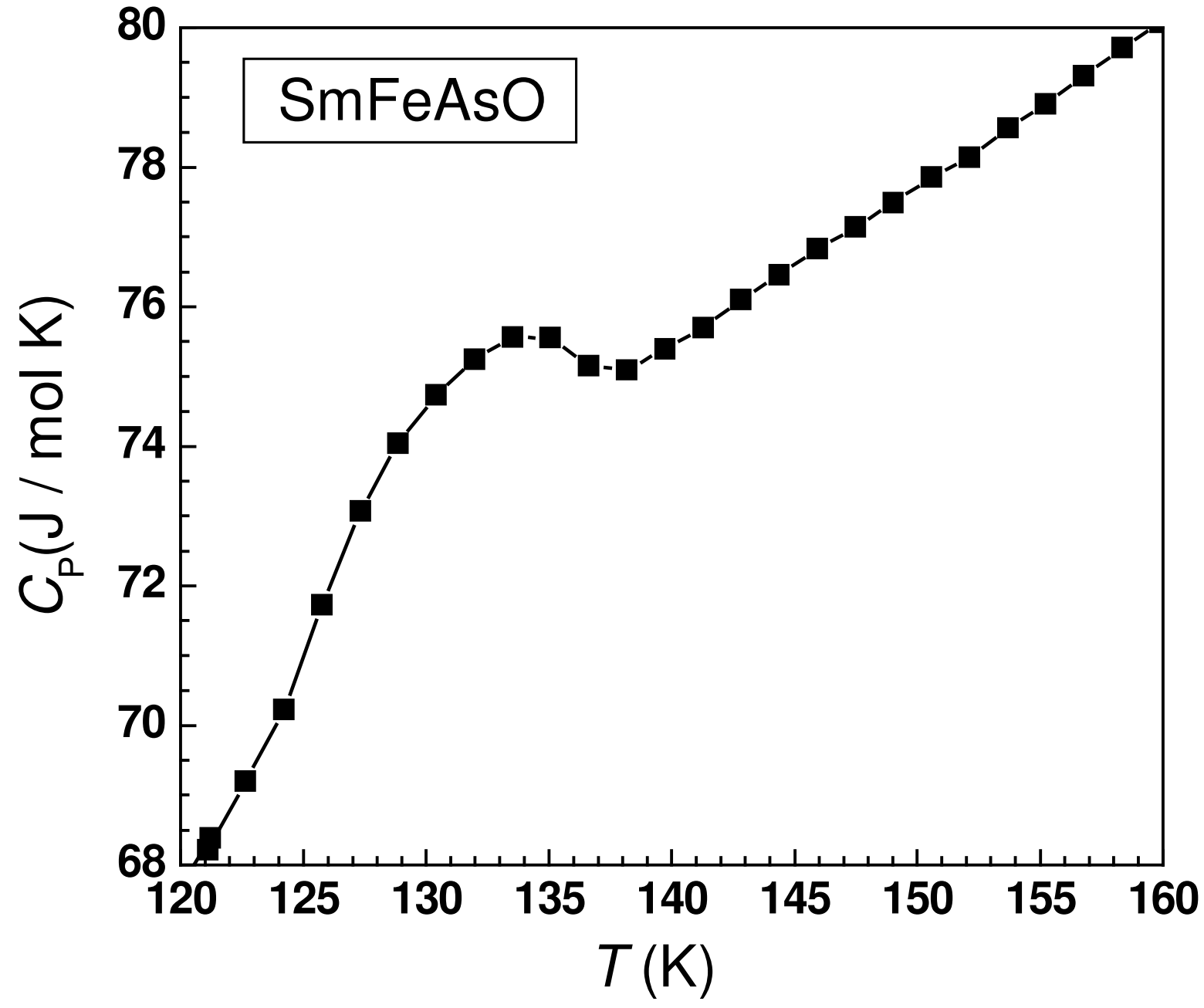

Figure 2 (b)

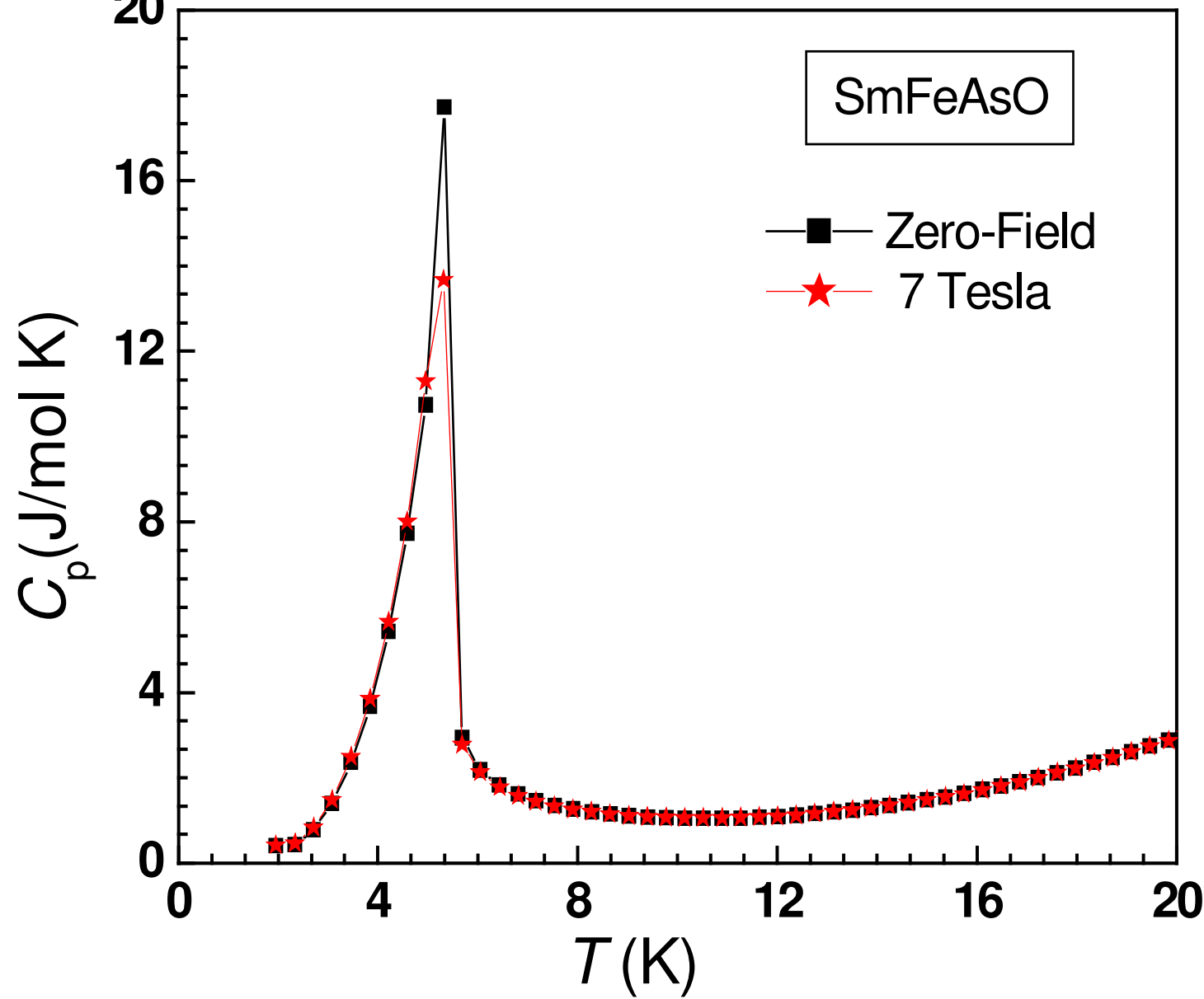

Figure 2 (c)

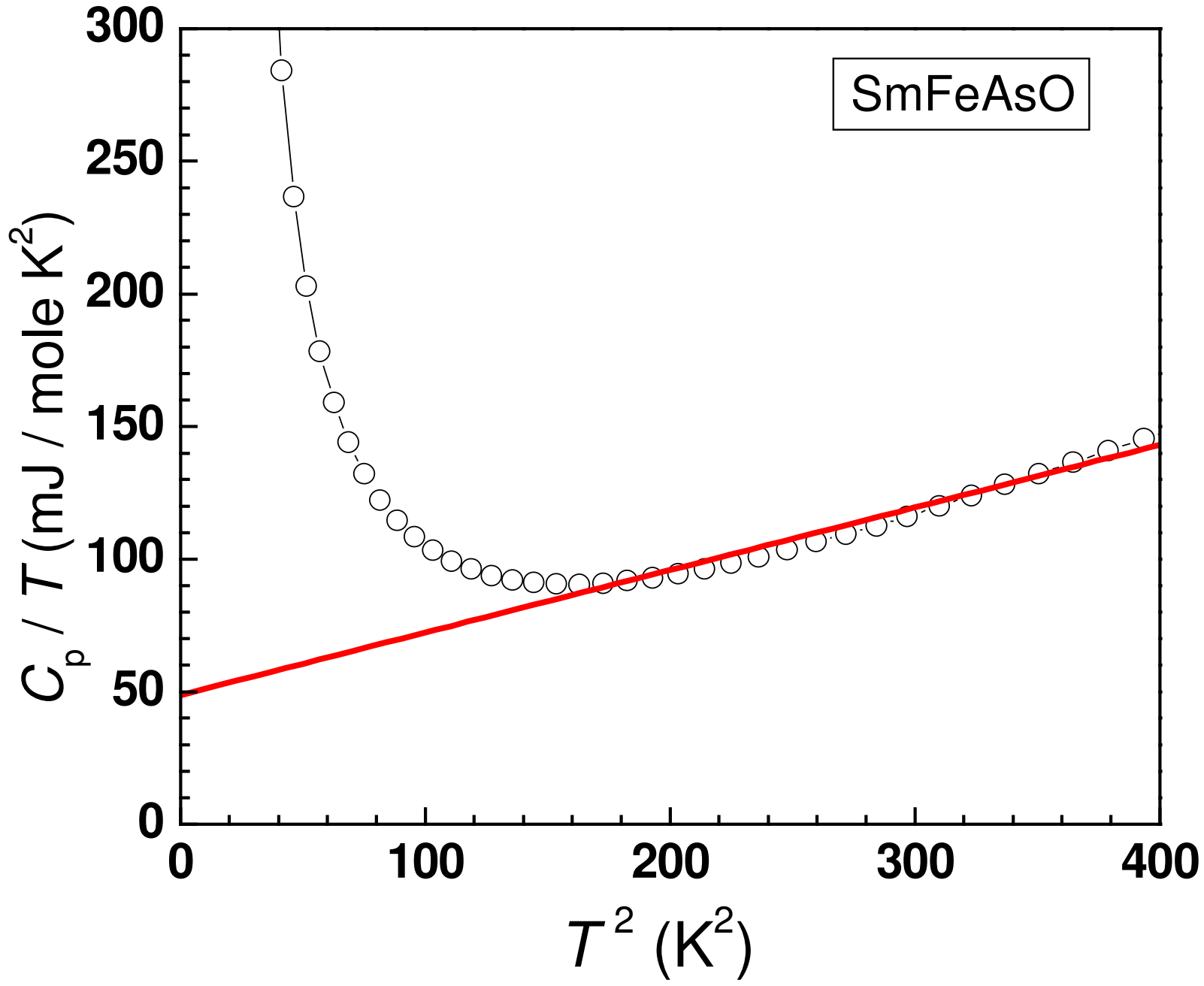

Figure 2 (d)

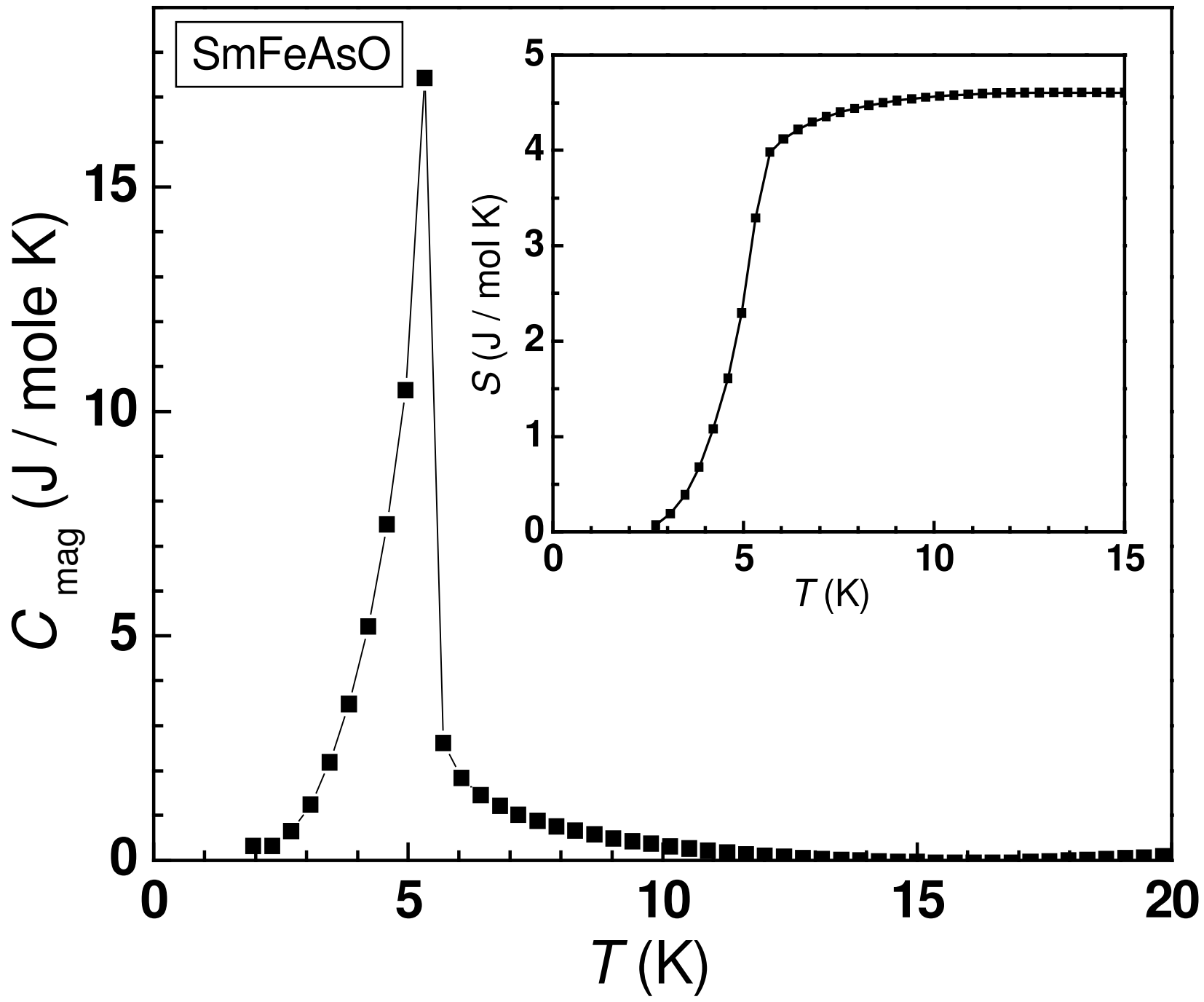